\documentclass[aapm,mph,reprint,graphicx]{revtex4-1}

\draft % marks overfull lines with a black rule on the right
\usepackage[english]{babel}
\usepackage[binary-units = true]{siunitx}
\usepackage{listings}
\usepackage{graphicx}
\usepackage{comment}
\usepackage{subfig}
\usepackage{threeparttable}
\usepackage{tikz}
\usepackage{tkz-euclide}
\usepackage{tikz-3dplot}
\usepackage{pgfplots}
\usepackage{filecontents}
\usepackage{tikzscale}
\usetikzlibrary{plotmarks}
\usepackage{grffile}
\usepackage{amsmath}
\usepackage{calc}
\usepackage{scalerel}				% used for reallywidetilde{}
\usepackage{stackengine,wasysym}		% used for reallywidetilde{}
\usepackage{amssymb} 				
\usepackage[outline]{contour}

\usetikzlibrary{shapes,arrows,arrows.meta,positioning,calc,decorations.markings,arrows.meta}

\newcommand{\quotes}[1]{``#1''}

\newcommand\reallywidetilde[1]{\ThisStyle{%
  \setbox0=\hbox{$\SavedStyle#1$}%
  \stackengine{-.1\LMpt}{$\SavedStyle#1$}{%
    \stretchto{\scaleto{\SavedStyle\mkern.2mu\AC}{.5150\wd0}}{.6\ht0}%
  }{O}{c}{F}{T}{S}%
}}

\begin{document}

%\preprint{}

%\title{Comprehensive Accuracy Evaluation\\of an Eye Tracking Based Eye Localization System} %Title of paper
%\title{Comprehensive Eye Localization Accuracy Evaluation for Proton Therapy} %Title of paper
%\title{Thorough Accuracy Evaluation\\of an Eye Tracking Based Eye Localization System} %Title of paper
%\title{Thorough Accuracy Evaluation of an Eye Tracker's Eye Location Estimate} %Title of paper
%\title{Eye Tracker Navigation Assessment:\\Evaluating the Accuracy of the Hidden Cornea Center} %Title of paper
%\title{Eye Tracker Evaluation:\\How Accurate is the Estimate of the Invisible Cornea Center?} %Title of paper
%\title{Eye Tracker Evaluation:\\How to Measure the Accuracy of the Invisible Cornea Center Location Estimate} %Title of paper
%\title{Navigation System for Eye Interventions:\\Accuracy Evaluation of the Invisible Eye Center Location\\ Estimated by an Eye Tracker} %Title of paper
%\title{Eye Tracker Accuracy Evaluation\\ of the Invisible Center of Corneal Curvature Location}
\title{Eye Tracker Accuracy: \\ Quantitative Evaluation of the Invisible Eye Center Location}



\author{Stephan Wyder}
\email[]{stephan.wyder@unibas.ch}
%\thanks{}
%\affiliation{University of Basel, Department of Biomedical Engineering, Allschwil, Switzerland}

\author{Philippe C. Cattin}
\email[]{philippe.cattin@unibas.ch}
%\thanks{}
\affiliation{Department of Biomedical Engineering, University of Basel, Allschwil, Switzerland}


\date{\today}
\begin{abstract}
\textbf{Purpose.}
%%%Introduction. In one sentence, what's the topic?
We present a new method to evaluate the accuracy of an eye tracker based eye localization system.
%Such a system can typically be used to navigate during a medical eye intervention, for instance in proton therapy.
%%%State the problem you tackle.
Measuring the accuracy of an eye tracker's primary intention, the estimated point of gaze, is usually done with volunteers and a set of fixation points used as ground truth.
However, verifying the accuracy of the location estimate of a volunteer's eye center in 3D space is not easily possible.
This is because the eye center is an intangible point hidden by the iris.
%%%Summarize (in one sentence) why nobody else has adequately answered the research question yet.
%An accuracy evaluation of the mentioned location estimate was not done so far to such an extent and with such realism.
%This kind of evaluation is not very often done, because most eye tracking applications are interested in the point of gaze accuracy, where the determined eye location is only a by-product.

\textbf{Methods.}
%%%Explain, in one sentence, how you tackled the research question.
We evaluate the eye location accuracy by using an eye phantom instead of eyes of volunteers.
For this, we developed a testing stage with a realistic artificial eye and a corresponding kinematic model, which we trained with $\mu \text{CT}$ data.
This enables us to precisely evaluate the eye location estimate of an eye tracker.

\textbf{Results.}
%%%In one sentence, how did you go about doing the research that follows from your big idea.
We show that the proposed testing stage with the corresponding kinematic model is suitable for such a validation.
Further, we evaluate a particular eye tracker based navigation system and show that this system is able to successfully determine the eye center with sub-millimeter accuracy.

\textbf{Conclusions.}
%%%As a single sentence, what's the key impact of your research?
We show the suitability of the evaluated eye tracker for eye interventions, using the proposed testing stage and the corresponding kinematic model.
The results further enable specific enhancement of the navigation system to potentially get even better results.
\end{abstract}


\maketitle %\maketitle must follow title, authors, abstract and \pacs

\section{Introduction}
%%%Move 1. Establishing a territory of research
Eye tracking devices, also known as eye- or gaze trackers are used to monitor eye movement.
An eye tracker is usually used to determine a person's point of gaze.
In market research, for instance, a wearable, video based eye tracking system can be used to uncover which product on which shelf is attracted by a test person.
Certainly, there exist other constructions of eye trackers (e.g. desktop or embedded devices) and many other eye tracking applications (e.g. in usability testing or in automotive industry) \cite{Narcizo:2013fq, Hansen:2010gq}.
Different physical principles might be behind an eye tracker, depending on the application \cite{Duchowski:2007dm}.
Video based eye trackers are the most widely used devices, because of their simplicity and the wide applicability.

In recent research, eye trackers are also used in navigation systems for computer assisted eye interventions \cite{Via:2015eu, Wyder:2016he, wyder_stereo_2016}.
In these cases, the eye tracker is used to estimate the 3D-location of the patient's eye, that is the eye center and orientation.
We define the eye center as the center of corneal curvature.
This can be useful to align an eye for an ophthalmic examination or treatment.
Furthermore, the point of gaze, estimated by the eye tracker, is automatically monitored to interrupt an examination or treatment in case of sudden eye motion.

%%%Step 1. Claiming centrality (47.5%)
Using an eye tracker for medical interventions demands high system accuracy.
This may decide between success or failure of an intervention because of the close proximity of critical structures within the eye.
For instance, an eye localization accuracy below $\SI{1}{mm}$ is required, when an eye tracker is used to target intraocular tumors.

The demand for accurate eye tracking systems also raises the need for reliable accuracy measurement methods.
Accuracy measurements are crucial for the development of an eye tracking system and also for the performance specification of the device.

%%%Step 2. Making topic generalisation (65.5%)
%The conventional eye tracker accuracy evaluation is done with volunteers, who have to focus on certain fixation points, which are located at well-known positions.
Conventionally, eye tracker accuracy is evaluated with volunteers, who have to focus on certain fixation points located at well-known positions.
The accuracy is then given by the deviations between the true fixation point locations and the point of gaze estimates of the eye tracker.
As straightforward as this evaluation can be performed on the one hand, as difficult it is to see what parts of the system contribute to a certain error on the other hand.
Testing this way does not enable us validating the accuracy of an intermediate product of the eye tracking pipeline, as for instance the eye center location.
Furthermore, this validation method obviously depends on the cooperation of the volunteers.
Hence, measuring the accuracy with an eye phantom seems to be the ideal complement for a thorough eye tracker evaluation.

%%%Step 3. Reviewing items of previous research (75%)
Already Via et al. \cite{Via:2015eu} used an eye phantom to asses the accuracy of an eye tracking system.
However, details about the exact procedure remain partially unclear.
Furthermore, it is not clear how realistic their eye phantom is.
Also \'{S}wirski and Dodgson recognized the lack of a comprehensive evaluation method to test and improve the individual parts of an eye tracking system.
They propose completely synthetic eye data \cite{Swirski:2014js} for accuracy and precision evaluation of eye tracking algorithms.

%%%Move 2. Establishing a niche
%%%Step 1a. Counter-claiming (disagreeing with others) (2.5%) 
Compared to Via et al. \cite{Via:2015eu}, we build up our ground truth data using $\mu \text{CT}$-measurements to get highly accurate absolute eye center locations.
In contrast to \'{S}wirski and Dodgson \cite{Swirski:2014js}, we do not only evaluate the algorithm, but we validate the complete eye tracking system, including the whole optical path and the external referencing to a medical device.
%We focus on the validation of the, for certain applications important, intermediate results of a video based eye tracker, the 3D eye location.
%%%Step 1b. Indicating a gap to be filled (57.5%)
The evaluation of such an eye localization system involving eye tracker hardware and its environment cannot be done with rendered eye images.
Neither can it be done with volunteer tests, because it is not possible to accurately measure the 3D location of a volunteers invisible eye center.

%%%Step 1c. Question-raising (22.5%)
%%%Step 1d. Continuing a research tradition (22.5%)

%%%Move 3. Occupying the niche
%%%Step 1a. Outlining purposes (defining research goal) (25%) 
Accurate ground truth data is required for the accuracy evaluation of an eye localization system.
We propose a procedure to fill this gap by providing accurate 3D-locations of the invisible and intangible eye center.

\begin{figure}
	\centering
	\includegraphics[width={246pt*0.95}]{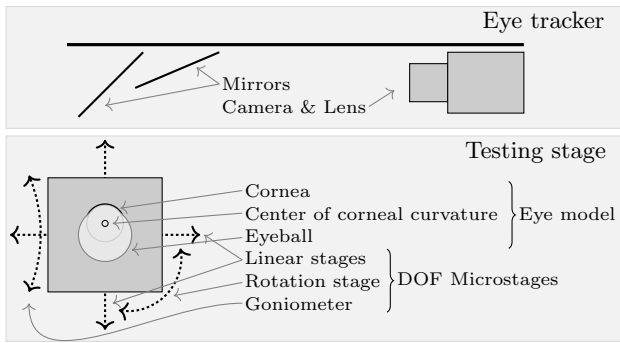}
 	\caption{Eye tracker and testing stage (topdown view)}
	\label{img:idea_sketch}
\end{figure}

%%%Step 1b. Announcing present research (95%)
The basis is formed by a testing stage with four degrees of freedom (4 DOF), a mounted artificial glass eye, and an attached, black and white checkerboard pattern for external referencing.
The testing stage enables us to move the whole eye forth and back and sidewards (by two linear stages).
Additionally, the testing stage enables us to rotate the eye around two axes (by a rotation stage and a goniometer), in order to simulate an arbitrary line of sight.
%We built and parametrized a corresponding kinematic testing stage model and trained it with $\mu \text{CT}$-data (i.e. high precision 3D volumetric data) acquired of the testing stage in several different configurations (i.e. eye positions and orientations).
We built the testing stage and trained the parameters of its kinematic model with $\mu \text{CT}$-data (i.e. high precision 3D volumetric data) acquired of the testing stage in several different configurations (i.e. eye positions and orientations).
The $\mu \text{CT}$-data provides us with accurate information about the location and the geometry of the eye and the checkerboard.
%Furthermore, we use the location information of the checkerboard pattern, acquired with the $\mu \text{CT}$ system, for external referencing.
Figure~\ref{img:idea_sketch} illustrates the two involved parts, the eye tracker we want to evaluate and the proposed testing stage to accomplish the evaluation.

%Having the testing stage ready and the kinematic model trained, the eye tracker is able to track the artificial eye to estimate the eye center location.
Having the testing stage ready and the kinematic model trained, we position and orient the artificial eye in known locations and compare this against the eye center locations predicted by the eye tracker.
%We collect the eye tracker estimates of the eye center locations for several different eye positions and orientations.
The eye center locations are given by the centers of corneal curvature (i.e. center of a cornea best fit sphere).
%To get the ground truth of this 3D-point, we further propose a corresponding testing stage model.
The trained kinematic model provides us with exactly the same point in a common coordinate system (CS), which is also accessible by the eye tracker.
This enables us to compare the eye location estimates of the eye tracker with the ground truth data, given by the testing stage model.

%%%Step 2. Announcing principal findings (70%)
%With the proposed testing stage, we can show that a particular eye tracking system \cite{wyder_stereo_2016} estimates the eye center location with sub-millimeter accuracy.
%Likewise, the proposed method can be used to evaluate the performance of any 3D model based eye tracker.
With the proposed testing stage, it is possible to quantitatively evaluate the performance of any 3D model based eye tracker.
Using this method, we show that a particular eye tracking system \cite{wyder_stereo_2016} estimates the eye center location with sub-millimeter accuracy.

%%%Step 3. Indicating research paper structure (70%)
We describe in the following sections our proposed method for the accuracy evaluation and the results achieved when testing a particular eye tracker \cite{wyder_stereo_2016}.
\section{Methods}
We propose a custom-built hardware testing stage and an appropriate kinematic model with its calibration, to evaluate the accuracy of the center location of the corneal curvature, estimated by an eye tracker.
This section consists of three parts.
First, we give an insight into a typical eye tracking model based on 3D ray tracing.
%This enables to understand what data our testing stage model has to provide, in order to be able to compare it to data provided by the eye tracker.
Second, we present the testing stage hardware with its components.
The testing stage hardware enables us to position and orient the embedded artificial eye such that the eye tracker can perform its intended measurements.
The testing stage hardware basically replaces the testing volunteer, with the advantage of having the exact position of the eye (i.e. ground truth).
Third, we present the corresponding testing stage model, which we parametrize, train and validate with $\mu \text{CT}$ data.
The kinematic model enables us to determine the exact glass eye position in every possible testing stage configuration with sub-millimeter accuracy.
%Without the kinematic model, we could test the eye tracker only with the exactly same microstage configurations, which we used for the $\mu \text{CT}$ scans.

Consequently, this enables us to test an eye tracker on the artificial eye prothesis with several different eye positions and orientations.
The 3D eye location estimate of the eye tracker can then be compared to the ground truth data of the testing stage model, which is configured according the status of the testing stage.
\subsection{Eye Tracker Model}

\begin{figure}
	\includegraphics{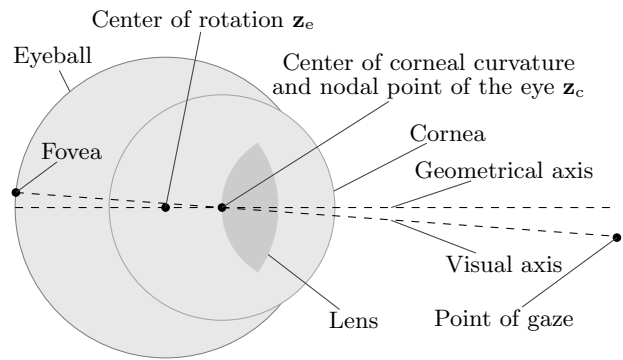}
 	\caption{Typical eye model used by eye trackers}
	\label{img:eye_tracker_model}
\end{figure}

The complexity of existing eye tracking models vary.
The eye is often modeled with two spheres.
Figure~\ref{img:eye_tracker_model} illustrates such a typical eye model.
One sphere represents the eyeball, its center consequently corresponds to the rotation center of the eye.
The second sphere, the sphere cap respectively, represents the cornea.
The center of the corneal curvature corresponds to the nodal point of the eye, where the optical rays cross, before they hit the retina \cite{Guestrin:2006dz}.

The two sphere centers define the geometrical axis of the eye.
Hence, the orientation of an eye in space can be determined by the geometrical axis.
The location of the eye (i.e. eye center) is given by the center of the corneal curvature, which lies on the mentioned, geometrical axis and is an integral part of most of the 3D model based eye trackers.

The fovea (point of sharpest vision) is located on the retina (backside of the eyeball) but is not in line with the geometrical axis.
A point we focus on with our eye gets imaged on the fovea.
That is why also the visual axis plays an important role in such a model.
The visual axis connects the fovea with the nodal point of the eye and the point of gaze.
The angle between visual axis and geometrical axis has to be calibrated per patient.%, depending on the eye tracking model with one or more calibration points.
%This, however, is usually done with a calibration procedure for each patient.
%With our testing stage, we focus on accuracy tests, which can not be tested with a patient or volunteer: the accuracy of the eye location estimate.

The eye tracker \cite{Wyder:2016he, wyder_stereo_2016} which we test with the proposed testing stage is based on the model of E. D. Guestrin and M. Eizenman \cite{Guestrin:2006dz}.
\subsection{Testing Stage Hardware}

\begin{figure}
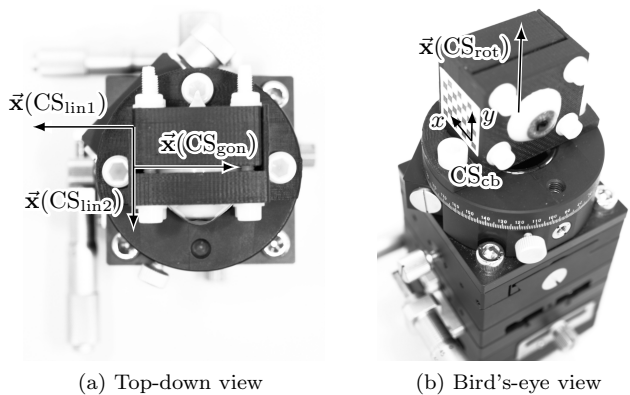

    \centering
    \subfloat[Top-down view]{\includegraphics[height={672pt*0.2}]{_TestingStage1.tikz} \label{fig:stage1}}
    \qquad
    \subfloat[Bird's-eye view]{\includegraphics[height={672pt*0.2}]{_TestingStage2.tikz} \label{fig:stage2}}
    \caption{Testing stage with the glass eye, its holder with the checkerboard and the microstages}
    \label{fig:testingstage1}
\end{figure}

The testing stage we developed consists of a translation stage with two axes with parameters $P_1$ and $P_2$ (\emph{OptoSigma TADC-652WS25-M6}), a goniometer stage with parameter $P_3$ (\emph{OptoSigma GOH-65A50-M6}), and a rotation stage with parameter $P_4$ (\emph{OptoSigma KSW-656-M6}).
The variables $P_1, P_2, P_3$, and $P_4$ represent the values, which are set for the corresponding microstages.
The linear stages have a vernier scale included enabling to measure with a precision of \SI{10}{\mu m}.
The rotation stage and goniometer also contain a vernier scale enabling us to measure with a precision of angular minutes.
To simulate the human eye, we use a handcrafted eye prosthesis made from glass by the \emph{Swiss Institute For Artificial Eyes, Lucerne, Switzerland}.
To interface the artificial eye with the stages we designed a rigid and robust eye holder.
Since the eye prosthesis does a priori not have an exactly known geometry, we made a 3D scan of it with a $\mu \text{CT}$ device (\emph{GE phoenix nanotom m}).
We segmented the surface of the eye with \emph{Fiji}, an image processing package \cite{Schindelin:2012ty}.
We afterwards used \emph{Blender}, an open source 3D creation suite (http://www.blender.org), to design a holder accurately interfacing the eye with the stages.
The holder additionally contains a black and white checkerboard stuck on its side.
The checkerboard is printed with an off-the-shelf laser printer, which contains toner visible in the $\mu \text{CT}$.
The stages are serially mounted and on top of them is the eye holder, which was printed on a \emph{Stratasys Fortus 250mc} 3D printer.
The testing stage is shown in Figure~\ref{fig:testingstage1}.
\subsection{Testing Stage Kinematic Model}
The aim of the kinematic model is to determine the exact center location of the corneal curvature $\mathbf{z}_\text{c}$ for a certain testing stage configuration ($P_1, P_2, P_3, P_4$) and to transform the coordinates to a common coordinate system.
%The common coordinate system corresponds to the checkerboard coordinate system, which is also accessible by the eye tracker.

\begin{figure}
	\centering
	\includegraphics{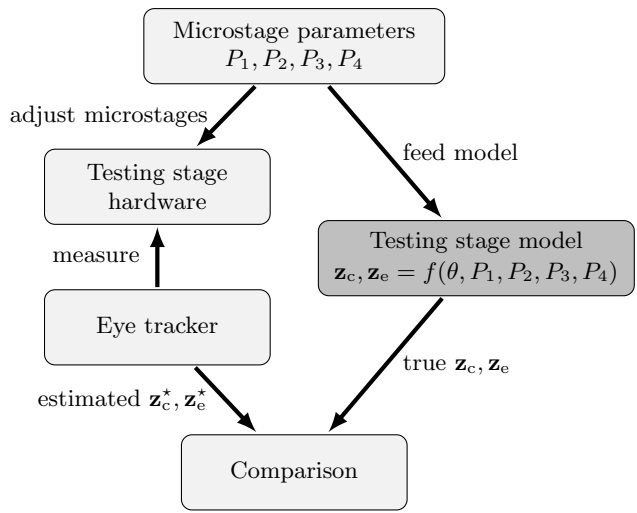}
 	\caption{Concept of testing stage: Comparison of the eye center location $\mathbf{z}_\text{c}^{\star}$ estimated by the eye tracker with the ground truth $\mathbf{z}_\text{c}$}
	\label{img:conceptTestingStage}
\end{figure}

The internal model parameters $\theta$, that have to be trained, basically consist of six right-handed coordinate systems (CS): $\text{CS}_\text{vol}$ is the common CS for all $\mu \text{CT}$ -volumes, $\text{CS}_\text{lin1}$, $\text{CS}_\text{lin2}$, $\text{CS}_\text{gon}$, and $\text{CS}_\text{rot}$ correspond to their appropriate microstage and $\text{CS}_\text{cb}$ is the checkerboard CS.
$\text{CS}_\text{vol}$ can be seen as the CS for model input data, whereas $\text{CS}_\text{cb}$ is the CS for the output data.
$\text{CS}_\text{cb}$ is accessible by the eye tracker and the testing stage.
Additionally, $\theta$ contains $\hat{\mathbf{z}}_\text{c}$ and $\hat{\mathbf{z}}_\text{e}$, the center locations and the radii of the cornea and the eyeball, yet unaffected by $P_1, P_2, P_3, P_4$ (neutral position).

Figure~\ref{img:conceptTestingStage} illustrates the role of the testing stage model within our contribution.

The origins of the CSs and the corresponding orientations are defined based on the acquired $\mu \text{CT}$ data.
We adjust a few positions of each individual microstage and acquire a $\mu \text{CT}$ volume for each configuration.
This enables us to train the internal kinematic model parameters $\theta$.

\textbf{$\boldsymbol\mu \text{CT}$  Data Acuisition.}
%\subsubsection{$\mu \text{CT}$  Data Acuisition}
As seen in Tab.~\ref{tab:dataacquisitionplan}, we acquired 15 $\mu \text{CT}$-volumes, which help us to define the mentioned internal model parameters $\theta$.
Furthermore, we used some $\mathbf{\mu} \text{CT}$ measurements to test the integrity of our kinematic model.
The table shows the number (identifier) of the measurement ($\#$), the state of the individual microstages during a certain scan and the type of the measurement ($\star$).

\begin{center}
\begin{threeparttable}
\caption{$\mu \text{CT-data}$ acquisition plan}
\label{tab:dataacquisitionplan}
\begin{tabular}{r|p{1.5cm}p{1.5cm}p{1.3cm}p{1.3cm}|l}
\hline \hline
\noalign{\vskip 1mm}
& \multicolumn{4}{c|}{Stages} & \\
& $P_1$ & $P_2$ & $P_3$ & $P_4$ &  \\
$\#$ & linear 1 & linear 2 & gonio. & rotation & $\star$ \\
\hline \noalign{\vskip 0.1cm}
1 & \SI{0}{mm} & \SI{0}{mm} & \SI{0}{\degree} & \SI{0}{\degree} & \scriptsize{1,3}\\
2 & \SI{-7.5}{mm} & \SI{0}{mm} & \SI{0}{\degree} & \SI{0}{\degree} &\scriptsize{3}\\
3 & \SI{+7.5}{mm} & \SI{0}{mm} & \SI{0}{\degree} & \SI{0}{\degree} &\scriptsize{5}\\
\hline \noalign{\vskip 0.1cm}
4 & \SI{0}{mm} & \SI{0}{mm} & \SI{0}{\degree} & \SI{0}{\degree} &\scriptsize{1,3}\\
5 & \SI{0}{mm} & \SI{-7.5}{mm} & \SI{0}{\degree} & \SI{0}{\degree}&\scriptsize{3} \\
6 & \SI{0}{mm} & \SI{+7.5}{mm} & \SI{0}{\degree} & \SI{0}{\degree}& \scriptsize{5}\\
\hline \noalign{\vskip 0.1cm}
7 & \SI{0}{mm} & \SI{0}{mm} & \SI{0}{\degree} & \SI{0}{\degree} &\scriptsize{1,4}\\
8 & \SI{0}{mm} & \SI{0}{mm} & \SI{-15}{\degree} & \SI{0}{\degree} &\scriptsize{4}\\
9 & \SI{0}{mm} & \SI{0}{mm} & \SI{+8}{\degree} & \SI{0}{\degree} &\scriptsize{5}\\
10 & \SI{0}{mm} & \SI{0}{mm} & \SI{+15}{\degree} & \SI{0}{\degree}& \scriptsize{4}\\
\hline \noalign{\vskip 0.1cm}
11 & \SI{0}{mm} & \SI{0}{mm} & \SI{0}{\degree} & \SI{0}{\degree} &\scriptsize{2,4}\\
12 & \SI{0}{mm} & \SI{0}{mm} & \SI{0}{\degree} & \SI{-30}{\degree}& \scriptsize{4}\\
13 & \SI{0}{mm} & \SI{0}{mm} & \SI{0}{\degree} & \SI{+15}{\degree} &\scriptsize{5}\\
14 & \SI{0}{mm} & \SI{0}{mm} & \SI{0}{\degree} & \SI{+30}{\degree}& \scriptsize{4}\\
\hline \noalign{\vskip 0.1cm}
15 & \SI{0}{mm} & \SI{0}{mm} & \SI{0}{\degree} & \SI{0}{\degree}&\scriptsize{2} \\
\noalign{\vskip 1mm}
\hline
\end{tabular}
\end{threeparttable}
\end{center}

$\star{\text{\scriptsize{1}}}$ corresponds to training scans, where the microstages are in neutral position.
For testing, we use $\star{\text{\scriptsize{2}}}$ scans, which also correspond to neutral position scans.
$\star{\text{\scriptsize{3}}}$ scans are used to train the linear stages.
$\star{\text{\scriptsize{4}}}$ scans are used to train the rotation and goniometer stages.
%$\star{\text{\scriptsize{5}}}$ scans are used to test the integrity of the degree of freedom of the individual microstages.
$\star{\text{\scriptsize{5}}}$ scans are used to test the kinematic model accuracy of the individual degrees of freedom.

To acquire the required data, we use the \emph{GE phoenix nanotom m} $\mu \text{CT}$ device.
In order to get a good contrast for the glass eye surface as well as for the checkerboard pattern in the acquired $\mu \text{CT}$ data, we set the voltage to \SI{50}{kV} and the current to \SI{310}{\mu A}.
To limit the required overall acquisition time for the 15 scans, we used a so called \emph{fast scan mode}, for which the specimen in the nanotom rotates continuously \SI{360}{\degree} during a defined time (in our case \SI{20}{\minute}).
These settings result in 1599 projections (\SI{3072 x 2400}{px}), exposed with \SI{750}{\milli \second} each.
The isotropic voxels have the side length of \SI{25}{\micro \meter}.
The resulting reconstructions (3D volumes) of the projections are cropped to the content of importance and have the size of \SI{2100 x 1900 x 1700}{px}.
Additionally, we reduce the grayscale depth from \SI{16}{\bit} to \SI{8}{\bit} by linearly mapping the grayscale-values between 23'000 and 35'000 to the range between 0 and 255, such that both, the eye surfaces as well as the checkerboards are well visible.
The whole process of reducing the volume dimensions and the grayscale depth is mainly required to reduce the amount of data for further processing.
The size of one final volume is still \SI{6,8}{\giga\byte}.

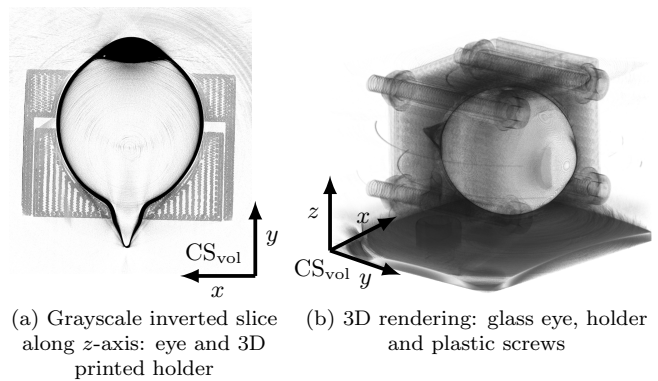
\begin{figure}
    \centering
    \subfloat[Grayscale inverted slice along $z$-axis: eye and 3D printed holder]{
    
    \begin{tikzpicture}
	\tikzstyle{every node}=[font=\small,style={inner sep=0,outer sep=1}]
	\node[anchor=south west,inner sep=0,transform shape] at (0,0) {\includegraphics[height={672pt*\real{0.15}}]{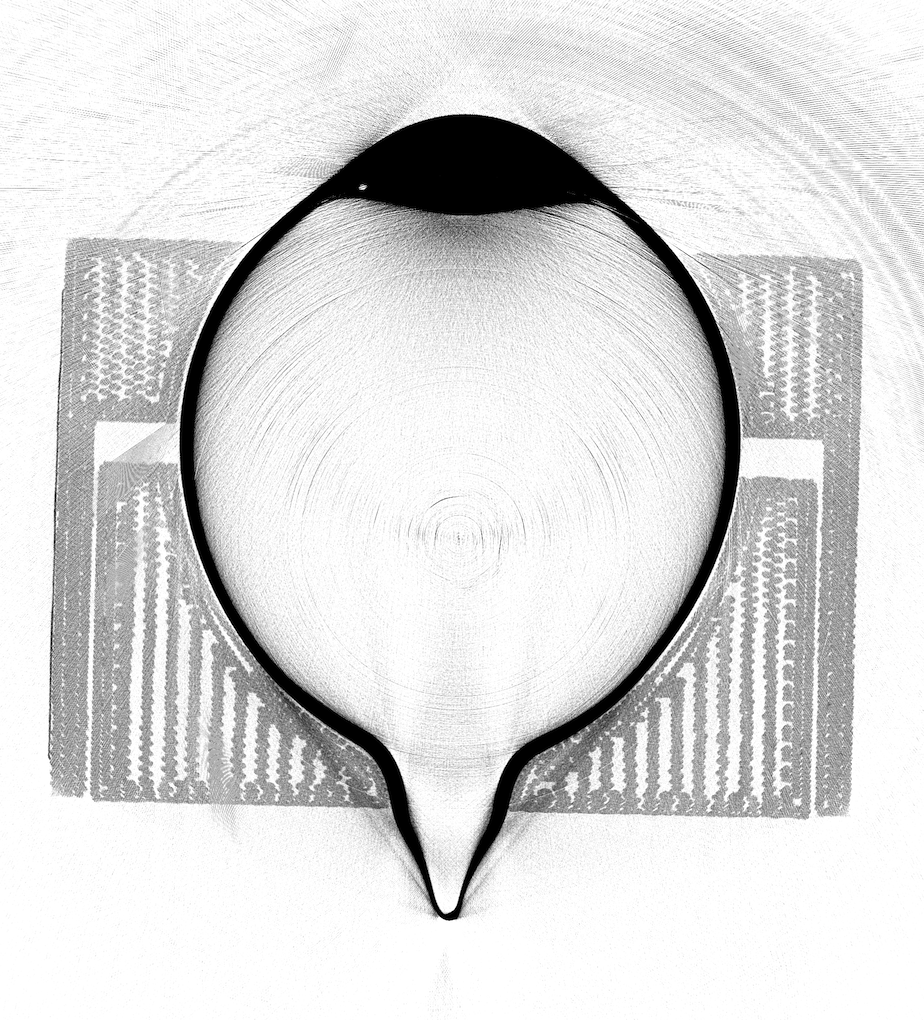}};
	%\tkzInit[xmax=3.5,ymax=3.5,xmin=0,ymin=0]
	%\tkzGrid
	%\tkzAxeXY
	\node at (2.75,0.3){$\text{CS}_\text{vol}$};
	\draw[-latex,line width=0.5mm] (3.3,0) -- (3.3,1) node[draw=none, midway, right=3pt]{$y$};
	\draw[-latex,line width=0.5mm] (3.3,0) -- (2.3,0) node[draw=none, midway, below=3pt]{$x$};
    \end{tikzpicture}
    
    \label{fig:uCT_Slice}}
   % \qquad
    \subfloat[3D rendering: glass eye, holder and plastic screws]{
    
    \begin{tikzpicture}
	\tikzstyle{every node}=[font=\small,style={inner sep=0,outer sep=1}]
	\node[anchor=south west,inner sep=0,transform shape] at (0,0) {\includegraphics[height={672pt*\real{0.15}}]{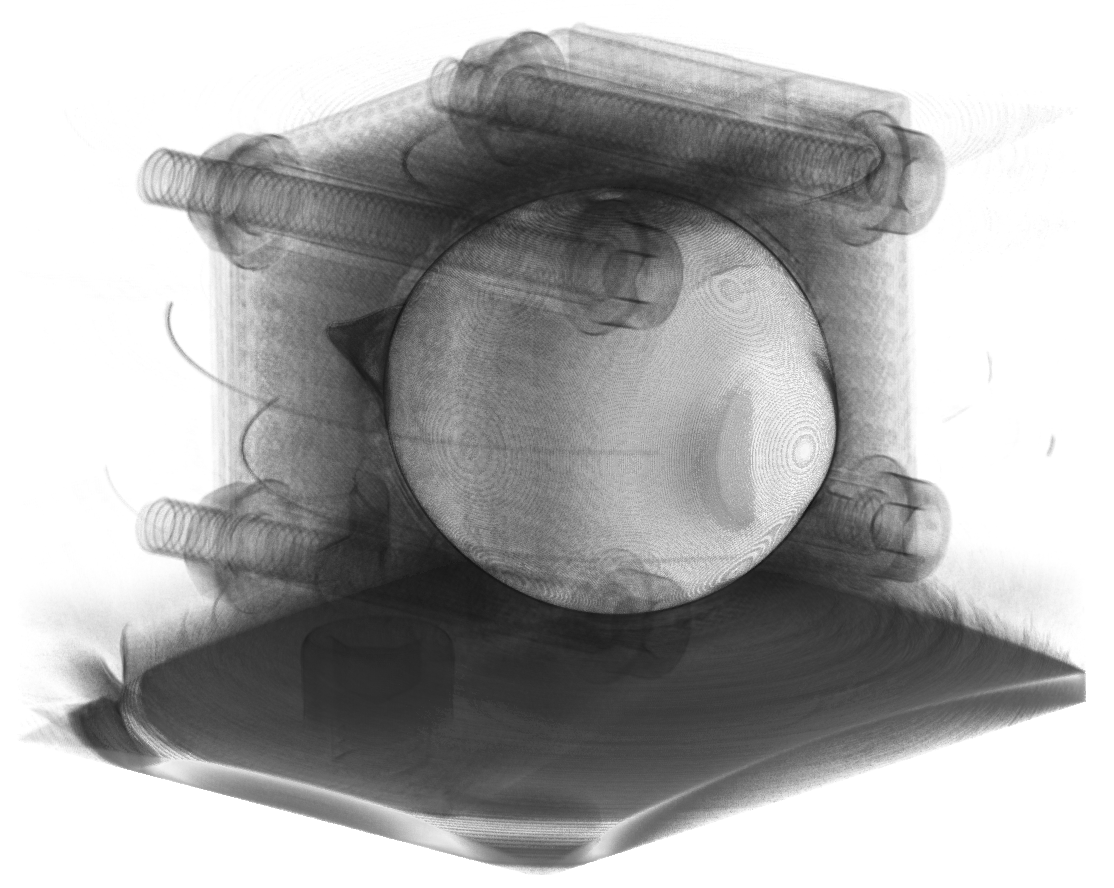}};
	%\tkzInit[xmax=4,ymax=4,xmin=0,ymin=0]
	%\tkzGrid
	%\tkzAxeXY
		
	\node at (00,0.4){$\text{CS}_\text{vol}$};
	\draw[-latex,line width=0.5mm] (0.1,0.65) -- (0.1,1.65) node[draw=none, midway, left=3pt]{$z$};
	\draw[-latex,line width=0.5mm] (0.1,0.65) -- +(-19:1cm) node[draw=none, midway, below=3pt]{$y$};
	\draw[-latex,line width=0.5mm] (0.1,0.65) -- +(27:1cm) node[draw=none, midway, above=3pt]{$x$};
    \end{tikzpicture}
    
    \label{fig:uCT_3D}}
    \caption{Visualized $\mu \text{CT}$ data ($\text{CS}_\text{vol}$) acquired with \emph{GE phoenix nanotom m}}
    \label{fig:uCTdata}
\end{figure}

Figure~\ref{fig:uCTdata} illustrates the data acquired with the $\mu \text{CT}$.
Figure~\ref{fig:uCT_Slice} shows one slice perpendicular to the $z$-axis and Figure~\ref{fig:uCT_3D} shows a volume rendering of a $\mu \text{CT}$ scan.
Both figures illustrate also the location and orientation of the $\text{CS}_\text{vol}$.

In order to be able to train our kinematic model with the acquired data, we first need to segment the required features.

\begin{figure}
	\includegraphics{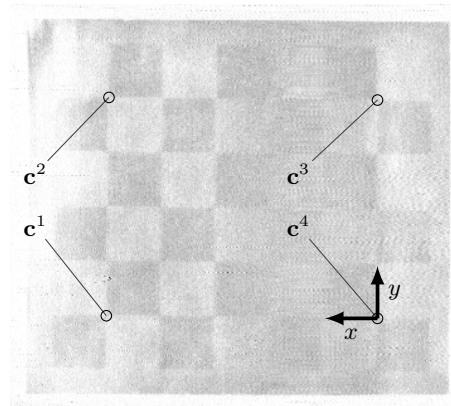}
 	\caption{Checkerboard corners ($\mathbf{c}^k$) and $\text{CS}_\text{cb}$ as seen in the $\mu \text{CT}$ data}
	\label{img:checkerboard}
\end{figure}

\textbf{$\boldsymbol\mu \text{CT}$ Data Segemention.}
%\subsubsection{$\mu \text{CT}$ Data Segemention}
We extract two different types of features from the acquired volumes, four checkerboard corners ($\mathbf{c}^k$, where $k \in \{1,2,3,4\}$), as they are visible in Figure~\ref{img:checkerboard}, and the surface of the glass eye (the black contour visible in Figure~\ref{fig:uCT_Slice}).

To train the kinematic model we need to have the corner point coordinates as they are visualized in Figure~\ref{img:checkerboard} for all 15 data volumes.
We extract the coordinates of $\mathbf{c}^k$ by hand using \emph{Fiji's} \quotes{Big Data Viewer}.
This plugin enables to visualize a slice with an arbitrary orientation and to show the 3D coordinates of a given voxel.
This results in $15*4$ 3D coordinates in $\text{CS}_\text{vol}$ coordinate system.

The following procedure describes the extraction of the glass eye for all volumes in neutral configuration ($\star{\text{\scriptsize{1}}}$ and $\star{\text{\scriptsize{2}}}$, see Tab.~\ref{tab:dataacquisitionplan}).
We process the volumes (thresholding and surface extraction) again by using \emph{Fiji} \cite{Schindelin:2012ty}.
The edge of the eye is segmented by applying a threshold of $115$, which is an experimentally found value.
Afterwards we extract the surface from the segmented eye using the marching cubes method (using the \quotes{3D Viewer} plugin).
The surface mesh can be exported as STL file directly with this plugin.
This results in a mesh basically consisting of an outer and an inner surface of the glass eye along with some unwanted holes and additional artifacts.

To clean up the geometry we import the mesh into \emph{Blender}.
Within \emph{Blender} we first create several objects by separating the imported mesh by loose parts.
All but the biggest part (the eye) can be deleted.
To save later processing time, we apply a mesh decimation.
%To get only the outer shell of the eye, we select everything visible and invert the selection. 
We extract the cornea and the eyeball separately to individually fit a sphere afterwards.
%To do so, we add two cubes to the scene.
%We adjust the size, align, and orient them roughly such that they contain the appropriate parts of the mesh.
%By using boolean modifiers, we can extract everything from the glass eye mesh, which is inside such a cube.
The cleaned cornea- and eyeball-mesh are exported again as STL for all 5 mentioned volumes.

After $\mu \text{CT}$ data acquisition and segmentation we end up with four 3D coordinates each ($\mathbf{c}^k$, $k \in \{1,2,3,4\}$) for all 15 volumes.
In addition we have an extracted cornea and an eyeball mesh for five of the 15 volumes (where $\star{\text{\scriptsize{1}}}$ and $\star{\text{\scriptsize{2}}}$).

\textbf{Kinematic Model Calibration.}
%\subsubsection{Kinematic Model Calibration}
All data used as input (checkerboard corner points, cornea mesh, eyeball mesh) to train the internal model parameters $\theta$ are in the right-handed $\text{CS}_\text{vol}$ coordinate system and are given in voxel.
We also express the other coordinate systems relative to $\text{CS}_\text{vol}$.

Let $\mathbf{c}^{k}_j \in \mathbb{R}^3$ be a 3D vector in $\text{CS}_\text{vol}$ representing a checkerboard corner point, where $k \in \{1,2,3,4\}$ encodes the checkerboard corner point number and $j \in \{1,2,3,...,15\}$ encodes the number of the measurement ($\#$).

Let $G^{k}_{p}$ be a group of $\mathbf{c}^{k}_j$, where $k \in \{1,2,3,4\}$ encodes the checkerboard corner point number and $p \in \{1,2,3,4,5\}$ encodes the type ($\star$) of the scan group (Tab.~\ref{tab:dataacquisitionplan}).

%Let $M_{q}$ be a measurement of four $\mathbf{c}^{k}_j$, representing all four corner points (CBC) of one measurement, where $q \in \{1,2,3,...15\}$ encodes the number of the measurement ($\#$).

A coordinate system is defined using four position vectors expressed in $\text{CS}_\text{vol}$.
The first column vector represents the origin $\vec{\mathbf{o}}$ of the corresponding CS expressed in $\text{CS}_\text{vol}$.
The remaining three column vectors represent the positions where the unit vectors (basis vectors) of the corresponding CS point to:
$$\text{CS} = \underbrace{
\left(\begin{array}{cccc}o_x & x_x &  y_x &  z_x \\
o_y & x_y & y_y  & z_y  \\
o_z & x_z & y_z & z_z \\1 & 1 & 1 & 1\end{array}\right)
}_{\substack{\text{Homogeneous coordinates in } \text{CS}_\text{vol} }}.$$
Usually a CS is represented with a rigid $4\times4$-transformation matrix (isometry) consisting of a rotation and a translation.
Our slightly different CS definition has the advantage, that the unit vectors can directly be extracted after a transformation is applied to the CS.

%or $j \in [\star1,\star5]$ encoding either the number(s) ($\#$) or the type(s) ($\star$) of the scan.

%$^{1}$= centerScanTrainData, $^{2}$= centerScanTestData, $^{3}$= linearStageVectors, $^{4}$= rotationStageVectors, $^{5}$= modelValidationData
%The remaining CSs ($\text{CS}_\text{lin1}$, $\text{CS}_\text{lin2}$, $\text{CS}_\text{gon}$, and $\text{CS}_\text{rot}$) correspond to the individual microstages.

First, we define $\text{CS}_\text{lin1}$ and $\text{CS}_\text{lin2}$, which represent the linear stage 1 and 2, the two stages at the bottom of the microstage stack.
The origins $\vec{\mathbf{o}}$ of $\text{CS}_\text{lin1}$ and $\text{CS}_\text{lin2}$ are given by the median ( $\widetilde{ }$ ) of three corner points, where $k=1$.
These three corner points come from volumes, where the stages were in neutral position during the scan ($\star{\text{\scriptsize{1}}}$ volumes):
$$\vec{\mathbf{o}}(\text{CS}_\text{lin1}) = \vec{\mathbf{o}}(\text{CS}_\text{lin2}) = \widetilde{G^1_{1}}.$$

The $x$-axes of $\text{CS}_\text{lin1}$ and $\text{CS}_\text{lin2}$ are pointing in the positive direction of the corresponding translational axis of the appropriate microstage.
They are defined using the median ( $\widetilde{ }$ ) of all four translation vectors %using all CBC's of $\star3$ volumes ($\#1,\#2,\#4,\#5$):
$$\vec{\mathbf{x}}(\text{CS}_\text{lin1}) = \vec{\mathbf{o}}(\text{CS}_\text{lin1}) + \frac{ \vec{\mathbf{x}_1} }{ \left\Vert \vec{\mathbf{x}_1} \right\Vert},$$
where $\vec{\mathbf{x}_1} = \reallywidetilde{\{\mathbf{c}^k_{1} - \mathbf{c}^k_{2} | k \in \{1,2,3,4\}\}}$ and
$$\vec{\mathbf{x}}(\text{CS}_\text{lin2}) = \vec{\mathbf{o}}(\text{CS}_\text{lin2}) + \frac{ \vec{\mathbf{x}_2} }{ \left\Vert \vec{\mathbf{x}_2} \right\Vert},$$
where $\vec{\mathbf{x}_2} = \reallywidetilde{\{\mathbf{c}^k_{4} - \mathbf{c}^k_{5} | k \in \{1,2,3,4\}\}}$.

The $y$-axis $\vec{\mathbf{y}}$ and $z$-axis $\vec{\mathbf{z}}$ of both systems are defined in an arbitrary way using the cross product, such that we get well defined right handed CSs with orthogonal axes.
Particular orientations of $\vec{\mathbf{y}}$ and $\vec{\mathbf{z}}$ are not important, since we use these two CSs only for translation along the $x$-axis.

Second, we define $\text{CS}_\text{gon}$ and $\text{CS}_\text{rot}$, which represent the goniometer and the rotation stages, the two topmost stages of the microstage stack.
The origins $\vec{\mathbf{o}}$ of $\text{CS}_\text{gon}$ and $\text{CS}_\text{rot}$ are given by best fit circle centers.
Because all checkerboard corners $k$ of the particular measurements lie in a plane perpendicular to the rotation axes of the stages, we take the median of the found circle centers.
To find the appropriate circle centers we fit for all four corner points $k$ a circle using three measurements per fit.
The best fit circle-function ($\text{BFC}$) \cite{Leon:1980vs} returns the center of the fitted circle:
$$\vec{\mathbf{o}}(\text{CS}_\text{gon}) = \reallywidetilde{\{\text{BFC}(\mathbf{c}^k_{7}, \mathbf{c}^k_{8}, \mathbf{c}^k_{10}) | k \in \{1,2,3,4\}\}},$$
$$\vec{\mathbf{o}}(\text{CS}_\text{rot}) = \reallywidetilde{\{\text{BFC}(\mathbf{c}^k_{11}, \mathbf{c}^k_{12}, \mathbf{c}^k_{14}) | k \in \{1,2,3,4\}\}}.$$

To define the $x$-axes (rotation axes) of the two topmost stages of the microstage stack, we take the normal vector perpendicular to the plane given by the appropriate corner points:
$$\vec{\mathbf{x}}(\text{CS}_\text{gon}) = \reallywidetilde{ \{(\mathbf{c}^k_{7} - \mathbf{c}^k_{8}) \times (\mathbf{c}^k_{7} - \mathbf{c}^k_{10})  | k \in \{1,2,3,4\}\}},$$ 
$$\vec{\mathbf{x}}(\text{CS}_\text{rot}) = \reallywidetilde{ \{(\mathbf{c}^k_{11} - \mathbf{c}^k_{12}) \times (\mathbf{c}^k_{11} - \mathbf{c}^k_{14})  | k \in \{1,2,3,4\}\}},$$
where $\times$ denotes the cross-product. 
The $y$-axis $\vec{\mathbf{y}}$ and $z$-axis $\vec{\mathbf{z}}$ of both systems are again defined in an arbitrary way using the cross product, such that we get well defined right handed CSs with orthogonal axes.

Third, we determine the center of the cornea best fit sphere, as well as the center of the eyeball best fit sphere based on the prepared mesh from measurement $\#1$.
To do so, we use the segmented and cleaned meshes and we fit a sphere in a least-square-sense \cite{Leon:1980vs}.
We first rearrange the general equation of a sphere,
$$(x_i-x_0)^2+(y_i-y_0)^2+(z_i-z_0)^2 = r^2,$$
such that we can write the expression in matrix notation and solve for the unknowns $x_0, y_0, z_0, \text{and } r$, which represent the center coordinates and the radius of the sphere.
The variables $x_i$, $y_i$, and $z_i$ are the coordinates of any point lying on the surface of the particular sphere.
This results in two vectors $\mathbf{z}_\text{c}$ for the cornea center and $\mathbf{z}_\text{e}$ for the eyeball center containing the best fit sphere center coordinates and the appropriate radius.

\begin{figure*}
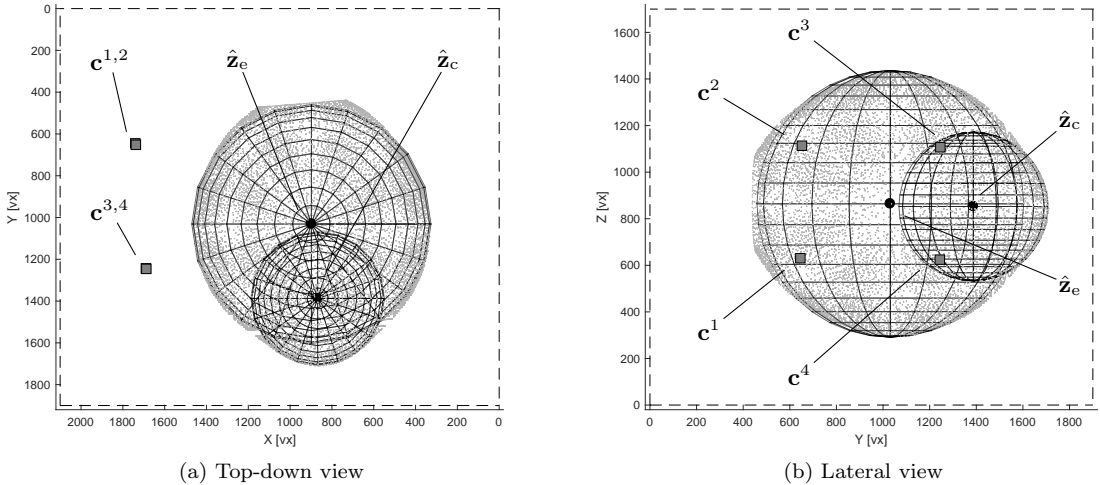

    \centering
    \subfloat[Top-down view]{\includegraphics[width={510pt * 0.4}]{_model_topdown.tikz} \label{fig:model1}}%
    \qquad
    \subfloat[Lateral view]{\includegraphics[width={510pt * 0.4}]{_model_side.tikz} \label{fig:model2}}%
    \caption{Testing stage model visualization}%
    \label{fig:model}
\end{figure*}

Figure~\ref{fig:model} illustrates $\mathbf{z}_\text{c}$, $\mathbf{z}_\text{e}$, and the vertices of the mesh (gray dots) with the corresponding best fit spheres (BFS).
The visualized checkerboard corners ($\mathbf{c}^k$) represent the median of the corners, where the stages are in neutral position($\star{\text{\scriptsize{1}}}$ containing \#1, \#4, and \#7).

The kinematic model is at this stage characterized such that we have defined four CSs corresponding to a microstage each and the centers and radii of the cornea and the eyeball.
All these position vectors are expressed in $\text{CS}_\text{vol}$.
In order to get the true position of the sphere centers (cornea or eyeball), we just have to translate $\mathbf{z}_\text{c}$ or $\mathbf{z}_\text{e}$ along the $x$-axis of linear stage CSs or rotate around the $x$-axis of the goniometer or rotation stage according to what is adjusted at the testing stage hardware (i.e. the microstages).

\textbf{Using the Kinematic Model.}
%\subsubsection{Using the Kinematic Model}
The trained testing stage model takes four parameters ($P_1,P_2,P_3,P_4$).
These are the four individual microstage position settings which are set on the testing stage hardware while the eye tracker estimates the cornea center for the corresponding eye position.
$P_1$ and $P_2$ are in millimeters (\SI{}{mm}).
$P_3$ and $P_4$ are in angular degrees (\SI{}{^\circ}).
Processing these parameters, the trained kinematic model is able to return (expressed in the common $\text{CS}_\text{cb}$) the position of the cornea center.
This position acts as ground truth for the eye tracker validation (Figure~\ref{img:conceptTestingStage}).
If we are adjusting a certain microstage position (e.g. $P_1= $ +\SI{6}{mm} on the linear stage 1), then this affects not only the position of $\mathbf{z}_\text{c}$ and $\mathbf{z}_\text{e}$, but also the microstages (their CSs, respectively) above the microstage which gets adjusted.
The microstage stack is as follows (from bottom to top): $\text{CS}_\text{lin1}$, $\text{CS}_\text{lin2}$, $\text{CS}_\text{gon}$, and $\text{CS}_\text{rot}$.
And on top of the stack is the eye with $\mathbf{z}_\text{c}$ and $\mathbf{z}_\text{e}$.

The workflow is as follows:
\begin{enumerate}
\item Hardware adjustment of a microstage $a$ ($a \in \{\text{lin1},\text{lin2},\text{gon},\text{rot}\}$)
\item Basis change from $\text{CS}_\text{vol}$ to the corresponding $\text{CS}_a$ of all remaining CSs, which are above the current $\text{CS}_a$ in the stack
\item Basis change to $\text{CS}_a$ of the sphere centers ($\mathbf{z}_\text{c}$ and $\mathbf{z}_\text{e}$)
\item Application of the transformation matrix $\text{T}_a$ (e.g. rotation of +\SI{3}{^\circ}) to all the remaining CSs and the sphere centers
\item Basis change of the CSs and the sphere centers back to $\text{CS}_\text{vol}$
\end{enumerate}
The workflow is repeated for all microstages (for all four parameters, respectively) beginning with the lowest one.

%The transformation from $\text{CS}_\text{vol}$ to $\text{CS}_{a}$ ($~^{a}\!T_\text{vol}$), correspond to what is set on the microstages.
The individual rigid transformations $\text{T}_a$, which are applied on the corresponding local CS look as follows (translation along or rotation around $x$-axis):
$$\text{T}_\text{lin1} = \left[\begin{array}{cccc}1 & 0 & 0 & P_1 \\0 & 1 & 0 & 0 \\0 & 0 & 1 & 0 \\0 & 0 & 0 & 1\end{array}\right], \text{T}_\text{lin2} = \left[\begin{array}{cccc}1 & 0 & 0 & P_2 \\0 & 1 & 0 & 0 \\0 & 0 & 1 & 0 \\0 & 0 & 0 & 1\end{array}\right],$$
$$\text{T}_\text{gon} = \left[\begin{array}{cccc}1 & 0 & 0 & 0 \\0 & \cos(P_3) & -\sin(P_3) & 0 \\0 & \sin(P_3) & \cos(P_3) & 0 \\0 & 0 & 0 & 1\end{array}\right],$$
$$\text{T}_\text{rot} = \left[\begin{array}{cccc}1 & 0 & 0 & 0 \\0 & \cos(P_4) & -\sin(P_4) & 0 \\0 & \sin(P_4) & \cos(P_4) & 0 \\0 & 0 & 0 & 1\end{array}\right].$$

The rigid transformations $~^{a}\!\text{T}_\text{vol}$ to change the basis from $\text{CS}_\text{vol}$ to $\text{CS}_a$ and back are defined as follows.
For this, we use a method based on singular value decomposition (SVD), which is robust in terms of noise \cite{Besl:1992iv}.
The method returns a rigid transformation $~^{a}\!\text{T}_\text{vol}$ (rotation and translation) when passing $\text{CS}_a$-matrix (expressed in $\text{CS}_\text{vol}$) and the $\text{CS}_\text{vol}$-matrix (expressen in $\text{CS}_\text{vol}$):
$$\text{CS}_\text{vol} = \underbrace{
\left(\begin{array}{cccc}0 & 1 & 0 & 0 \\0 & 0 & 1 & 0 \\0 & 0 & 0 & 1 \\1 & 1 & 1 & 1\end{array}\right)
}_{\substack{\text{Homogeneous coordinates in } \text{CS}_\text{vol} }}.$$
Having $~^{a}\!\text{T}_\text{vol}$, we change the basis of the remaining CSs (CSs above the current one in the microstage stack), $\mathbf{z}_\text{c}$, and $\mathbf{z}_\text{e}$.
Afterwards, we apply the transformation $\text{T}_a$ and change the basis back to $\text{CS}_\text{vol}$ for all CSs $b$, which are above $\text{CS}_a$:
$$\text{CS}_b' =  (~^{a}\!\text{T}_\text{vol}~)^{-1}\cdot(~\text{T}_{a}\cdot(~^{a}\!\text{T}_\text{vol} \cdot \text{CS}_b~)).$$
where $a$ represents the CS, which we adjust (e.g. $\text{CS}_\text{lin1}$).
The sphere centers are adjusted as well for each parameter $P_1,P_2,P_3,P_4$:
$$\mathbf{z}_\text{e}' =  (~^a\!\text{T}_\text{vol}~)^{-1}\cdot(~\text{T}_a\cdot(~^a\!\text{T}_\text{vol} \cdot \mathbf{z}_\text{e}~)),$$
$$\mathbf{z}_\text{c}' =  (~^a\!\text{T}_\text{vol}~)^{-1}\cdot(~\text{T}_a\cdot(~^a\!\text{T}_\text{vol} \cdot \mathbf{z}_\text{c}~)).$$
Step-by-step, we apply all transformations for a certain testing stage configuration, until we have the position $\mathbf{z}_\text{c}$ and $\mathbf{z}_\text{e}$ for the current microstage configuration expressed in $\text{CS}_\text{vol}$.
The last step is to change the basis of the sphere centers from $\text{CS}_\text{vol}$ to $\text{CS}_\text{cb}$, our common CS.
%$\text{CS}_\text{cb}$ corresponds to the right-handed checkerboard coordinate system, which is accessible by the eye tracker and the testing stage model.

For the eye tracker tests, the tracker is rigidly mounted to a certain position, such that the checkerboard pattern (also attached to the eye holder) is completely visible by the eye tracker camera.
For the external referencing of the eye tracker (here with the testing stage) we perform a homography estimation \cite{Zhang:2000fu} based on a checkerboard pattern \cite{Wyder:2016he, wyder_stereo_2016}.
This enables the eye tracker to express its guess about the sphere centers in $\text{CS}_\text{cb}$.
We configure the testing stage (adjusting linear, rotation, and goniometer stages) such that the visibility of the checkerboard pattern from the eye tracker is well (sharp and complete pattern).
This particular stage configuration enables us to access $\text{CS}_\text{cb}$ from our kinematic model.
The origin lies on the corner point 4, the $x$-axis points towards corner point 1 and the $y$-axis points towards corner point 3  (Figure~\ref{img:checkerboard}).
This $\text{CS}_\text{cb}$ definition holds for both the eye tracker and the testing stage model.

The workflow described above is applied again at the very end to transform the sphere centers to $\text{CS}_\text{cb}$ according to the microstage configuration ($P_1$, $P_2$, $P_3$, $P_4$) at the time of external referencing.
\section{Experiments}
\subsection{Kinematic Model consistency}
To make sure that we trained our testing stage model sufficiently accurate, we used the $\mu \text{CT}$ measurements of type $\star{\text{\scriptsize{5}}}$ and $\star{\text{\scriptsize{2}}}$ (see Tab.~\ref{tab:dataacquisitionplan}) to validate the integrity of the trained $x$-axes of the individual CSs.
We used the median checkerboard corner points of the measurements $\star{\text{\scriptsize{2}}}$ ($\{ \widetilde{G^1_{2}},\widetilde{G^2_{2}},\widetilde{G^3_{2}},\widetilde{G^4_{2}}\}$) to predict with our testing stage model the new checkerboard corner locations under four certain configurations.
We used one configuration ($P_1,P_2,P_3,P_4$) for each DOF.
For this, we took the four different configuration sets from the measurements $\star{\text{\scriptsize{5}}}$.
Having the new checkerboard corner locations calculated, we compared the model estimates (based on measurements $\star{\text{\scriptsize{2}}}$) with the checkerboard corners, which we extracted manually (measurements $\star{\text{\scriptsize{5}}}$).
The mean error (corner-reprojection-error) of the four $\star{\text{\scriptsize{5}}}$-measurements times four checkerboard corners (16 points) was \SI{31}{\micro\meter}.

Additionally, we analyzed the angles between the $x$-axes of the trained coordinate systems ($\text{CS}_{lin1}$, $\text{CS}_{lin2}$, $\text{CS}_{gon}$, and $\text{CS}_{rot}$).
Assuming the microstages are ideally mounted and aligned on top of each other, we would have to expect angles of $\SI{90}{^\circ}$ between the $x$-axes.
We found out that we have a $\SI{89.5}{^\circ}$ angle between the linear stages, $\SI{90.9}{^\circ}$ between the linear stage 2 and the goniometer rotation axis and $\SI{90.2}{^\circ}$ between the rotation axes of the goniometer and the rotation stage.

We also performed cornea-fit-refit experiments, where we fitted a new sphere to all of the scans $\star{\text{\scriptsize{5}}}$.
The mean deviation between the five sphere centers was $\pm$ \SI{36}{\micro\meter}.
\subsection{Eye tracker accuracy}
\textbf{Setup.}
%\subsubsection{Setup}
We tested a video based stereo eye tracker \cite{wyder_stereo_2016} with the proposed testing stage hardware and the corresponding kinematic model.
For this, we rigidly mounted both the testing stage and the eye tracker on an optical bench and aligned the eye tracker such that a good visibility on to the artificial eye of the testing stage was given.
We adjusted the focus and the aperture of the lens (part of the eye tracker) and performed a camera calibration \cite{Zhang:2000fu} to get the intrinsic camera parameters (focal length, distortions).
Having the camera calibrated, we adjusted the testing stage such that the holder's checkerboard was visible by the eye tracker
($P_1 = +\SI{8}{mm},P_2 = +\SI{7}{mm},P_3 = \SI{8}{^\circ},P_4 = +\SI{56}{^\circ}$).
With the eye tracker we performed a homography estimation (based on an image snapshot of the checkerboard) in order to be able to transform the eye tracker output, the center of the corneal curvature, to the common checkerboard coordinate system $\text{CS}_\text{cb}$ \cite{Wyder:2016he}.
The camera calibration and the referencing to an external system (testing stage or a medical device) is part of the eye tracker calibration procedure.

For the actual validation, we set 20 different eye positions and orientations with the testing stage to mimic snapshots of a natural eye movement.
To get a better impression of the results we only adjusted one microstage at the same time, while the three other stages were in neutral position.
The microstages were set to $P1 = \{7.5,10,12.5,15,17.5\}$[mm], then $P2 = \{7.5,10,12.5,15,17.5\}$[mm], $P3 = \{-10,-5,0,5,10\}$[$^\circ$], and $P4 = \{290,298,307,316,324\}$[$^\circ$].
This resulted in five positions per microstage and with that in 20 eye tracker estimates of the corneal curvature location $\mathbf{z}_\text{c}^{\star}$.
We set the same parameters on our kinematic model and generated the ground truth of the center location of the corneal curvature.
Figure~\ref{img:conceptTestingStage} illustrates this workflow.

\begin{figure}
	\centering
	\includegraphics[]{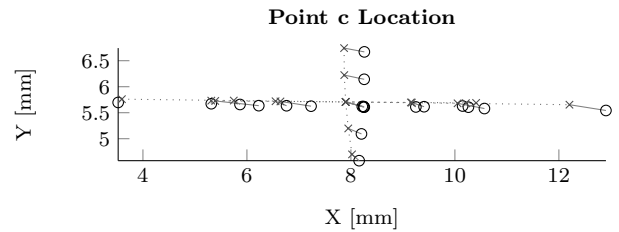}
 	\caption{Accuracy in the $X$/$Y$ plane (o = eye tracker estimate, x = ground truth, $...$ = DOF)}
	\label{img:res1}
\end{figure}

\begin{figure}
	\centering
	\includegraphics[]{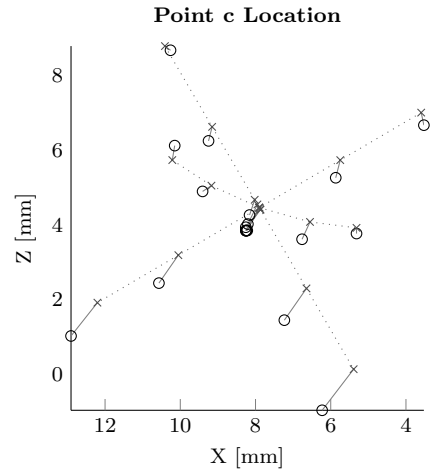}
 	\caption{Accuracy in the $X$/$Z$ plane (o = eye tracker estimate, x = ground truth, $...$ = DOF)}
	\label{img:res2}
\end{figure}

\textbf{Results.}
%\subsubsection{Results}
We compared the 20 different center locations of corneal curvature from the eye tracker with the ground truth data from the testing stage.
The mean deviation between two 3D points, the accuracy $a$ respectively is as follows:
The mean accuracy $\mu(a) = \SI{0.68}{mm}$, the median accuracy $\widetilde{a} = \SI{0.67}{mm}$.
Subdivided into the individual orientation components:
The mean accuracy $\mu(a_x) = \SI{0.32}{mm}$, the median accuracy $\widetilde{a_x} = \SI{0.33}{mm}$.
The mean accuracy $\mu(a_y) = \SI{-0.09}{mm}$, the median accuracy $\widetilde{a_y}= \SI{-0.09}{mm}$.
The mean accuracy $\mu(a_z) = \SI{-0.54}{mm}$, the median accuracy $\widetilde{a_z} = \SI{-0.55}{mm}$.
Figure~\ref{img:res1} and Figure~\ref{img:res2} illustrate the distribution of the error.

Thanks to the proposed method we were able to analyze the nature of the error and unveil a slight bias of a yet unknown source.
For this, we removed the average error vector from our eye tracker estimates and compared the result again with the ground truth, then we got a mean relative error $\mu(a_{\text{rel}}) = \SI{0.32}{mm}$.
By eliminating this error, the overall eye tracker accuracy can even be increased.

We also evaluated the accuracy of the eye orientation.
For this, we calculated the geometrical axes for the eye tracker estimate by using the pupil center and the center of corneal curvature $\mathbf{z}_c^\star$ and for the kinematic model by using the centers of both spheres $\mathbf{z}_c$, $\mathbf{z}_e$.
In theory, all four points lie on the geometrical axis, however, it is not the case for our eye phantom.
That is why we calculated the relative angle between the geometrical axes from one measurement to the next and then we compared these relative angles between the ground truth and the eye tracker estimates.
The mean relative angle error is \SI{0.50}{\degree}, which indicates high angular precision.

\newlength\figurewidth
\setlength\figurewidth{246pt * \real{0.75}}

\section{Discussion}
We were able to successfully validate the eye tracker of interest with our testing stage hardware and the corresponding kinematic model.
The tests showed that the eye tracker can determine the eye location (center of corneal curvature) with an accuracy below \SI{0.7}{mm}.
The accuracy of the validated navigation system for proton radiotherapy hence fulfills the requirements of sub-millimeter accuracy.
The mean relative error $\mu(a_{\text{rel}})$ is smaller by roughly a factor of two compared to the mean error $\mu(a)$, which is a strong indication for high precision but also for a slight bias of a yet unknown source.
Our system helped to detect and quantify this bias.

Figure~\ref{img:res1} and Figure~\ref{img:res2} show this slight systematic error along the longitudinal axis of the eye tracker.
%This remaining error may have several reasons and may originate from both the testing stage (kinematic model) and the eye tracker.

It is difficult to compare the results to any other similar validation method, because to our best knowledge, no one did so far such a comprehensive validation of the eye location accuracy.
Having for instance a closer look at \cite{Via:2015eu}, it is not clear how exactly the ground truth was generated.

\subsection{Testing stage hardware and kinematic model}
The systematic error from the eye tracking tests may be explained by an imprecise cornea best fit sphere.
We prepared the cornea mesh in a way, where we only had limited influence on the vertex distribution.
Fitting a sphere with another method than with a least-square method might be more accurate.

%In any case, with a voxel size of \SI{0.025}{mm}, we can not measure more precisely than with $\pm \SI{0.025}{mm}$.
%It might be worth decreasing the voxel size.

Maybe the most important error source is the manual segmentation of the checkerboard corner points.
To improve this, we suggest exchanging the checkerboard pattern, which is used on the one hand for the external referencing of the eye tracker (homography) and on the other hand to train and validate the whole testing stage model.
Hence, the pattern, its segmentation respectively, is central for the validation.
A better pattern might be dots in a certain arrangement (similar to the squares in the checkerboard pattern).
This pattern could easily be segmented automatically, by choosing the center of mass of the circles or the ellipsoids, respectively, taking the thickness of the ink into account.

\subsection{Eye tracker}
Depending on the application different levels of accuracy are required. 
Our achieved sub-millimeter accuracy in determining the eye location is sufficient for our medical application with especially high demands.
If there should be higher demands, the detailed validation results, for instance the distribution of the error, might provide helpful information for eye tracker improvement.

\section{Conclusion}
Using an eye tracker to localize the eye in space can potentially improve today's eye interventions.
For instance, when treating eye tumors with protons, our non-invasive eye tracker based solution might some day replace the state-of-the-art invasive navigation method.

We proposed a quantitative evaluation method with which we showed that our eye tracker is able to fulfill the requirements, namely, to determine the location of the eye with sub-millimeter accuracy.
Our proposed evaluation method does not replace the eye tracker tests with volunteers that are used nowadays, but it complements the validation, enabling new eye tracking applications: eye localization.

We are sure, that in the future more and more applications, especially in ophthalmology, will benefit from an eye localization system.

%
%
%
%
%
%
%
%

% If in two-column mode, this environment will change to single-column format so that long equations can be displayed. 
%\begin{widetext}
%$$\mbox{put long equation here}$$
%\end{widetext}

% Figures should be put into the text as floats. 
% Use the graphics or graphicx packages (distributed with LaTeX2e).
% Put the label that you will use with \ref{} command in the braces of the \label{} command.
%
% \begin{figure}
% \end{figure}

% \begin{table}
% \end{tabular}
% \end{table}

\vspace{0.8cm} % manually addded
\begin{acknowledgments}
We would like to thank the members of the \emph{Biomaterials Science Center (BMC)} of the University of Basel for their support with the data acquisition with the $\mu \text{CT}$ system.
We thank also Otto E. Martin from the \emph{Swiss Institute For Artificial Eyes} for providing us a hand-crafted eye and for sharing his profound knowledge.
The work is funded by the \emph{Swiss National Science Foundation (SNSF)}.
\end{acknowledgments}

\bibliographystyle{ama}
\bibliography{bibliography}

\twocolumngrid
\printtables
\printfigures

\end{document}